\documentclass[aps,pra,twocolumn,amsmath,amssymb,nofootinbib,showpacs]{revtex4-1}
\usepackage[english]{babel}
\usepackage{latexsym}
\usepackage{graphics}
\usepackage{graphicx}
\usepackage{epsfig}
\usepackage{color}
\usepackage{bm}
\usepackage{amsmath}
\usepackage{amssymb}
\usepackage{amsthm}
\usepackage{dcolumn}
\usepackage{bm}
\usepackage{float}
\usepackage{hyperref}
\usepackage{color}
\usepackage{epstopdf}
\usepackage{cleveref}
\usepackage[svgnames]{xcolor}
\hypersetup{hidelinks,colorlinks=true,allcolors=DarkBlue}

\begin{document}

\preprint{APS/123-QED}

\title{Educational potential of quantum cryptography \\ and its experimental modular realization}

\author{A.K.~Fedorov}
\affiliation{Russian Quantum Center, Skolkovo, Moscow 143025, Russia}
\author{A.A.~Kanapin}
\affiliation{Russian Quantum Center, Skolkovo, Moscow 143025, Russia}
\author{V.L.~Kurochkin}
\affiliation{Russian Quantum Center, Skolkovo, Moscow 143025, Russia}
\author{Y.V.~Kurochkin}
\affiliation{Russian Quantum Center, Skolkovo, Moscow 143025, Russia}
\author{A.V.~Losev}
\affiliation{Russian Quantum Center, Skolkovo, Moscow 143025, Russia}
\author{A.V.~Miller}
\affiliation{Russian Quantum Center, Skolkovo, Moscow 143025, Russia}
\author{I.O.~Pashinskiy}
\affiliation{Russian Quantum Center, Skolkovo, Moscow 143025, Russia}
\author{V.E.~Rodimin}
\affiliation{Russian Quantum Center, Skolkovo, Moscow 143025, Russia}
\author{A.S.~Sokolov}
\affiliation{Russian Quantum Center, Skolkovo, Moscow 143025, Russia}
\date{\today}
\begin{abstract}
The fundamental principles of quantum mechanics are considered to be hard for understanding by unprepared listeners, 
many attempts of its popularization turned out to be either difficult to grasp or incorrect. 
We propose quantum cryptography as a very effective tool for quantum physics introduction as it has the desired property set to intrigue students and outline the basic quantum principles. 
A modular desktop quantum cryptography setup that can be used for both educational and research purposes is presented. 
The carried out laboratory and field tests demonstrated usability and reliability of the developed system.
\end{abstract}
\maketitle

We live in a time of dynamically developing quantum technologies.\,Over the last decades an essential breakthrough happened in new applied aspects of quantum physics, 
such as secure communication networks, 
sensitive sensors for biomedical imaging and fundamentally new paradigms of computation. 
While not wholly agreeing on whether this is the second or third quantum revolution~\cite{Dowling,Sowa}, 
the worldwide scientific society points out that the new emerging quantum technologies now promise the next generation of products with exciting 
and astounding properties that will affect our lives profoundly. 
In an effort to accelerate the impact of these innovations, many quantum manifests, memorandums, strategies, 
and other roadmaps were drawn up and adopted in recent years~\cite{Heracleous,Manifesto,National}. 
The importance of education of a new generation of technicians, engineers, scientists and application developers in quantum technologies is emphasized everywhere. 

However, we assume that there is arguably very poor awareness of quantum physics amidst non-physicists --- even worse than that of GTR. 
Most of well-educated people are able to say something about space deformation near a massive object, 
whereas quantum mechanics does not have a similar key phrase which is easy to recall. 
It is all the more strange that every day we use the achievements of quantum technologies like semiconductors, but hardly deal with relativistic phenomena. 
That general ignorance of quantum physics sometimes leads to mishaps, 
such as the clueless reaction of world media to the recognition by the Russian authorities the prospects of quantum teleportation~\cite{Oliphant} in June, 2016. 
It was reported by many media sources that Russian scientists were going to develop teleportation, as it was understood in science fiction. 
All in all, it is important to popularize quantum physics, spreading information about new technologies, achievements and challenges, creating the right social and regulatory context.

\begin{figure}[t]
\begin{center}
\includegraphics[width=1\columnwidth]{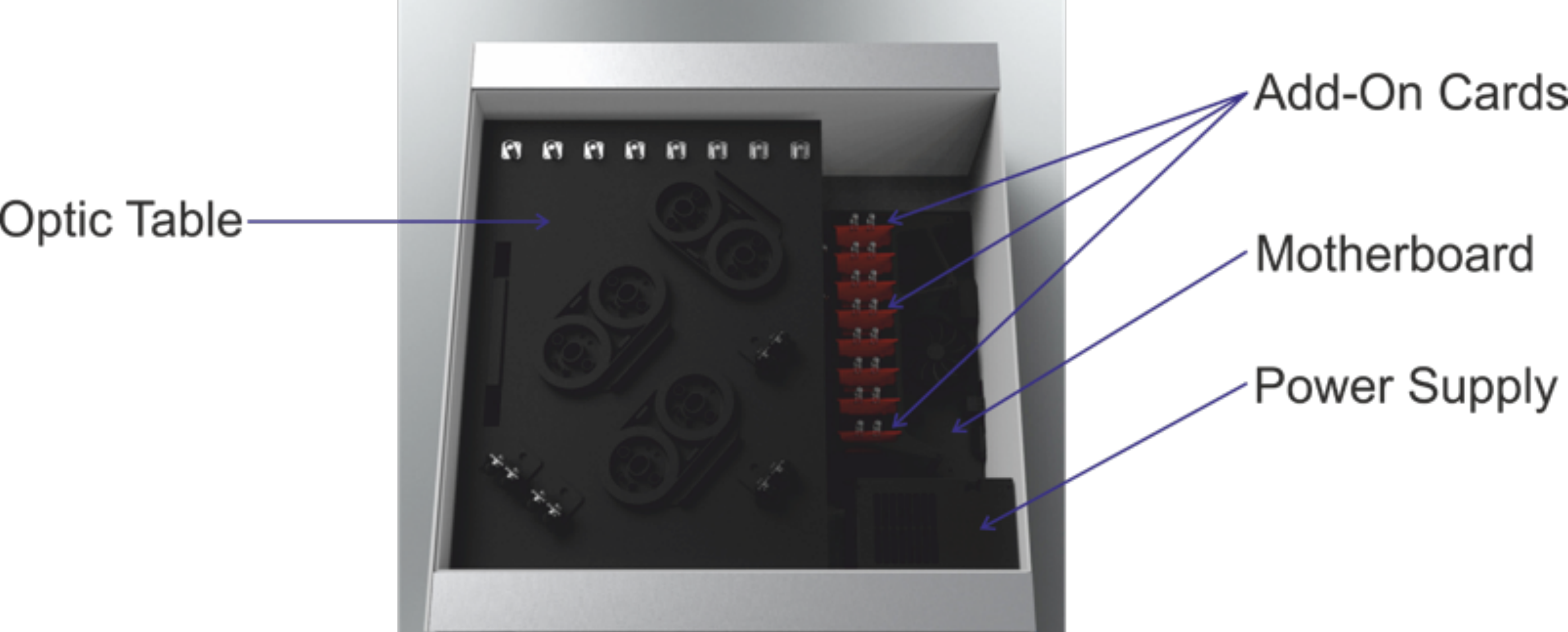}
\vskip -2mm
\caption
{Design concept of the Alice's unit in the developed modular QKD setup.}
\label{fig:modular}
\end{center}
\end{figure}

Meanwhile there are certain difficulties in quantum physics popularization related to the very nature of empirical science. 
Many phenomena on the quantum scale have no analogues in the common human world so they cannot be expressed via our everyday language. 
Some spectacular and animated attempts to make an interpretation of quantum mechanics by means of metaphysics or 
Eastern mysticism are rejected out of hand by the great majority of physicists as distorting reality~\cite{Dieks}. 
Besides, most physicists have no clear conception of the interpretation of their most basic theory, quantum mechanics. 
As a result, a common introduction to quantum mechanics starts with the definition of the mathematical formalism and its interpretation, 
which is difficult to grasp for an unprepared listener and is not suitable for popularization. 
An important tool in popularization consists in making scientific results exciting. 
The best way to intrigue students is to make use of descriptions that are both easy to understand and unusual~\cite{Dieks}.

We consider quantum cryptography to be a very suitable subject for an introductory role to quantum mechanics. 
Indeed, it allows to grasp the very fundamental principles of quantum physics, 
such as superposition, non-orthogonal quantum states and so on, actively involving all the necessary concepts to begin with. 
Secondly, it is rather easy to understand. 
According to our experience it can be basically explained to high school students within half an hour without extraordinary mental efforts. 
Thirdly, it is mysterious and exciting because, on one hand, associates with cryptic spy games, and on the other, it guarantees protection based on the laws of nature. 

\begin{figure}[t]
\begin{center}
\includegraphics[width=1\columnwidth]{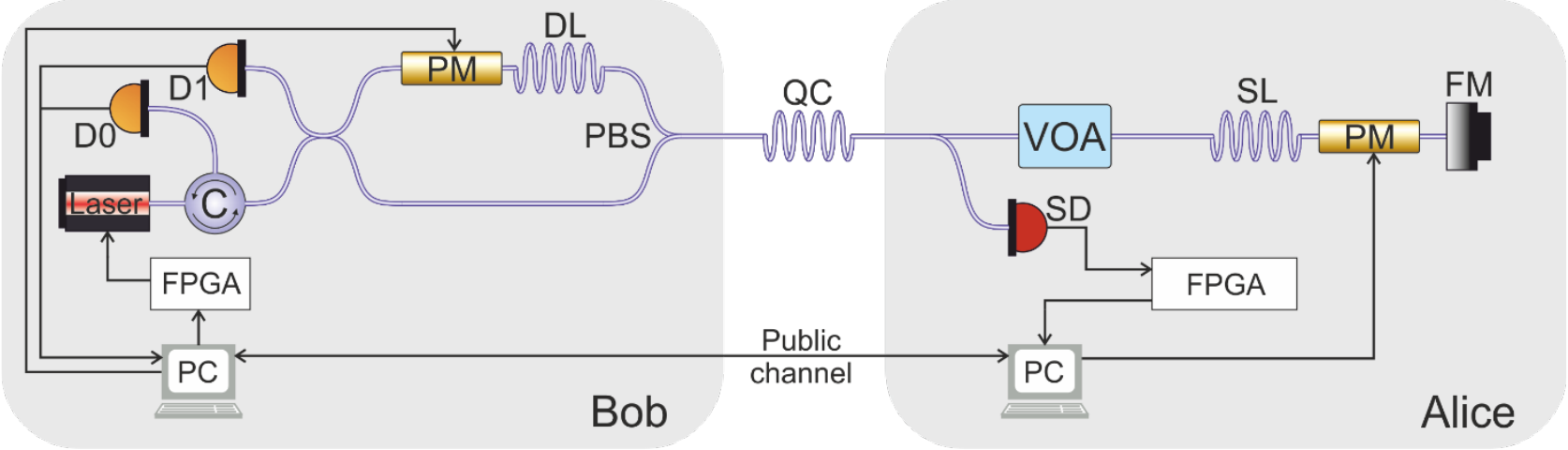}
\vskip -2mm
\caption
{Schematic diagram of the QKD system 
Laser is the semiconductor laser, 
C is the circulator, 
DL is the delay line, 
PM is the phase modulator, PBS is the polarization beam-splitter, 
D0 and D1 are single photon detectors, 
FM is the Faraday mirror, 
SL is the storage line, 
QC is the quantum channel, 
VOA is the variable optical attenuator, 
SD is the synchronization detector, 
and PC it the personal computer.}
\label{fig:modular}
\end{center}
\end{figure}

Considering quantum cryptography with respect to the methodological potential, its inspiring aspect deserves to be mentioned. 
There are several fundamental negative assertions in quantum physics: 
impossibility of making any measurement without perturbation of the system, the Heisenberg uncertainty principle, and the no-cloning theorem. 
Quantum cryptography has a very advantageous look on this background, not prohibiting but allowing absolutely secure cryptographic key distribution.

Let us briefly outline the problems of quantum cryptography and quantum communication in general. 
Secure communication is important for a wide range of consumers, including enterprises and governments. 
However, the methods used at present are based on classical cryptography, which can be broken with the completion of a quantum computer. 
Thus, post-quantum cryptography methods are in need of development, since they cannot be broken by quantum computers. 
Solutions are already commercially available today, as are tools for quantum random number generation, which are a key foundation in most cryptographic protocols. 
One of the problems is that they are limited by distance. Existing quantum key distribution devices can operate only over distances up to 300km. 
Such a weakness arises from quantum cryptography's own strength – since information cannot be cloned, it also cannot be relayed through conventional repeaters. 
To reach global distances, trusted nodes or quantum repeaters, perhaps even involving satellites, are needed. 
Trusted nodes allow for lawful intercept, which is required by many states, whereas quantum repeaters can extend the distance between trusted nodes, 
taking advantage of multimode quantum memories. 
The technologies for fully quantum repeater schemes have been tested in laboratory conditions, and are only several years away from reaching the market. 
Although long-distance transmission is only possible via photons, quantum memory can be realized via trapped ions, atoms in optical resonators, quantum dots and so on, 
allowing storage and processing at repeater nodes. 
Companies, such as ID Quantique, Toshiba and British Telecom, are being increasingly involved with quantum communication technologies.

Quantum cryptography is a relatively young science. 
One can say that it was born in the '70s as an idea that belonged to Stephen Wiesner who proposed ``quantum money'', which cannot be forged by copying. 
In the '80s it was developed as a first quantum key distribution (QKD) protocol BB84~\cite{BB84} and got its first experimental implementation. 
Whilst the experiment in the '90s was considered by a majority of physicists mainly as an expensive toy of philosophical nature, nowadays it has industrial realizations. 

We propose a QKD system specialized for student education purposes and scientific research. 
It has a module structure and can be modified easily. 
The QKD system is based on the LabVIEW platform and for control signals National Instruments (NI) FPGA boards are used. 
We use this platform because of that tool's high popularity in research activity of the academic community, to which our product is aimed for. 
On the basis of our system, numerous student workshops can be carried out in universities. 
Due to the quantum basis and optical methods of information transfer, this desktop device suits pretty well for the education of undergraduate physics students. 
Taking into account challenging tasks to generate, control, modify and synchronize electric and optical pulses, it is a powerful practical course for electronics engineers and hardware programmers. Involving the necessity to correct errors in sifted quantum keys and privacy amplification, we get a wide field for research for algorithm developers and mathematicians. 
In sum, a QKD workshop system is an indispensable instrument for general training. 
All these areas are naturally included in the new promising engineering field --- photonics. 
The photonics industry shows dynamic growth these years, the world photonics market is estimated to be EUR 615 billion by 2020~\cite{Photonics21}. 
The Russian government adopted a roadmap for the photonics development in 2013~\cite{PhotonicsRoadmap}.

The proposed desktop QKD system ``University Apparatus'' traditionally consists of two parts Alice and Bob, 
the former one is in charge of sending quantum signals, the last one is for receiving them. 
Fig.~1 illustrates a design concept of the Alice unit, Bob looks identical. 
On the right side, one can see a row of 8 red add-on electronic cards with SMA connectors. 
The laser module, phase and amplitude modulator drivers are implemented in the form of such cards. 
These cards can be inserted into the motherboard, situated on bottom of the unit. 
The main function of the motherboard is to provide commutation between add-on cards and the NI board installed in an ordinary PC (not shown in the figure) 
and connected to the motherboard via a PCI cable. 
In addition, the motherboard implements the power supply control and executes a delaying of electric pulses to apply them at the correct time to optic modulators. 
On the top side of the Bob or Alice unit there is an optic table where fiber optic components can be settled. 
On the optic table fiber component trays and mating sleeves are shown in the figure. 
The parts that do not require easy access are located under the optic table. 
Thereby we have a system, which is easily reachable from all sides, where we can reconFig.~both the optical scheme and electronic driver cards set. 
A sufficient amount of additional SMA connectors allows one to connect an oscilloscope for learning the shape of electrical pulses and their position on the timescale. 
The units are supplied with removable covers and side handles for transportation.

Using a similar modular design, the plug\&play QKD scheme has been assembled and tested in different circumstances (see Fig.~2). 
Among the many possible QKD optical schemes, the plug\&play one is characterized by relatively simple configuration and parameters adjustments, 
so it is ideal to begin the experimental study of quantum cryptography. 
The plug\&play scheme consists of a two-pass auto compensation fiber optic scheme with Mach-Zehnder interferometric phase shift measurements. 
A detailed description of the plug\&play scheme operation principles has been given in many sources, for example see Ref.~\cite{Bouwmeester}. 
The semiconductor laser LDI-DFB2.5G under control of FPGA board Spartan-6 generates optical pulses 2 ns wide on the standard telecommunication wavelength 1.55 $\mu$m 
with a frequency of 10 MHz. 
The circulator, beam-splitters, Faraday mirror and phase modulators are quite standard components, we used ones by Thorlabs. 
The quantum channel and storage line are single mode fibers each 25km long. 
For the synchronization of electric and optic pulses in Alice's unit, either a detector, as in the original plug\&play system, or a direct coaxial cable leading from Bob to Alice was used. 
The synchronization coaxial cable may be utilized in an academic QKD system for its simplification and cost reducing. 
Artix-7 FPGA board is used for the delay of synchronization electric pulses. 
The single photon detectors are free-running ID Quantique ID230, whose count registers only in 5 ns time windows when a photon can be expected. 
Use as short as possible time windows can effectively reduce the quantum bit error rate (QBER) level. 
Overall control of the electro-optic components is accomplished by NI PCIe-7811R installed in the corresponding PC --- Bob or Alice. 
Key distribution procedure comprises of repetitive sessions of 1000 trains, with 2400 pulses in each train. 
The raw key is being sifted after each session via the public channel by TCP/IP protocol.

\begin{figure}[t]
\begin{center}
\includegraphics[width=1\columnwidth]{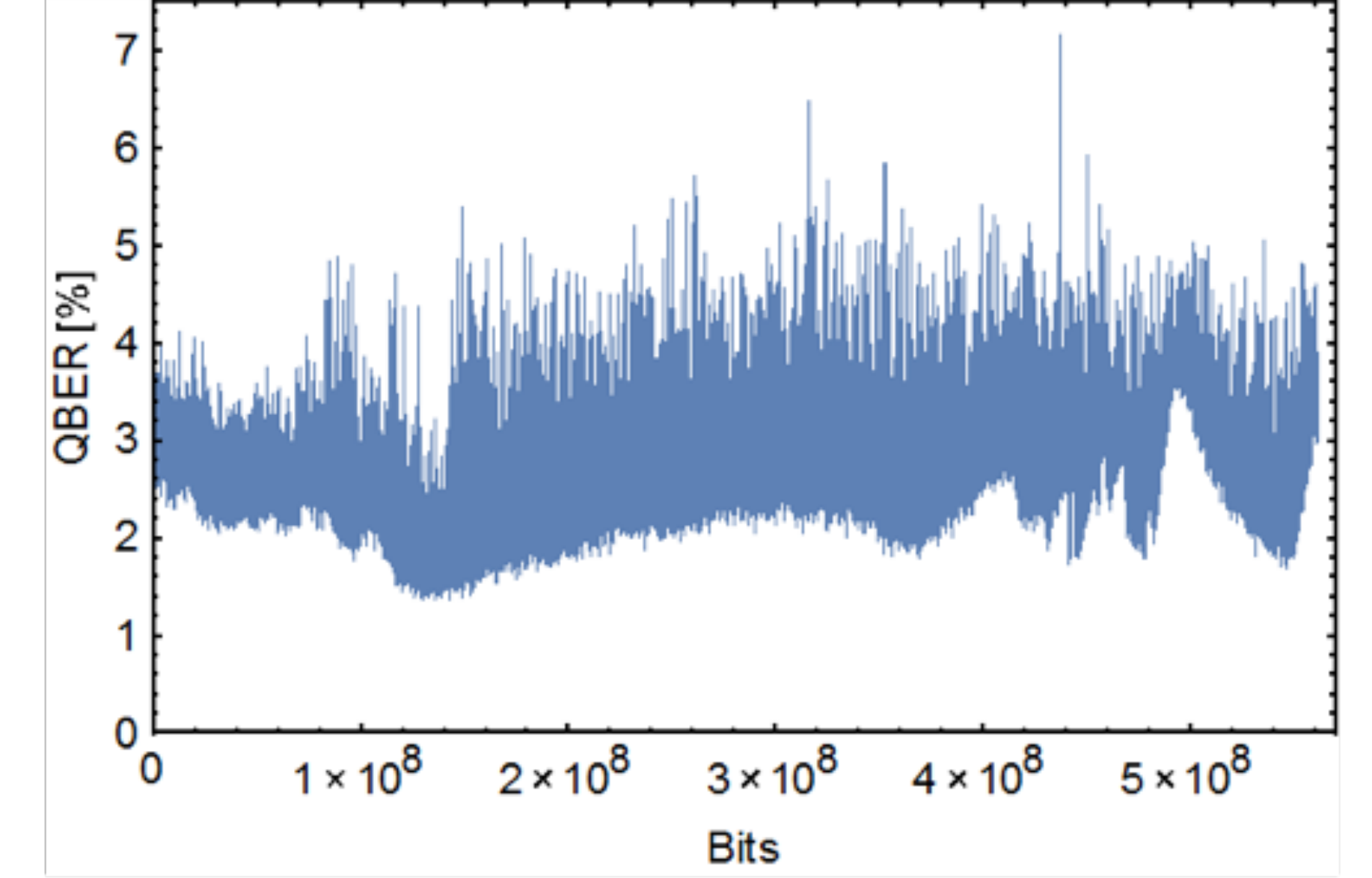}
\vskip -3mm
\caption
{QBER level is shown as function of distributed key size at five-day continuous experiment.}
\label{fig:modular}
\end{center}
\end{figure}

\begin{figure}[h]
\begin{center}
\includegraphics[width=1\columnwidth]{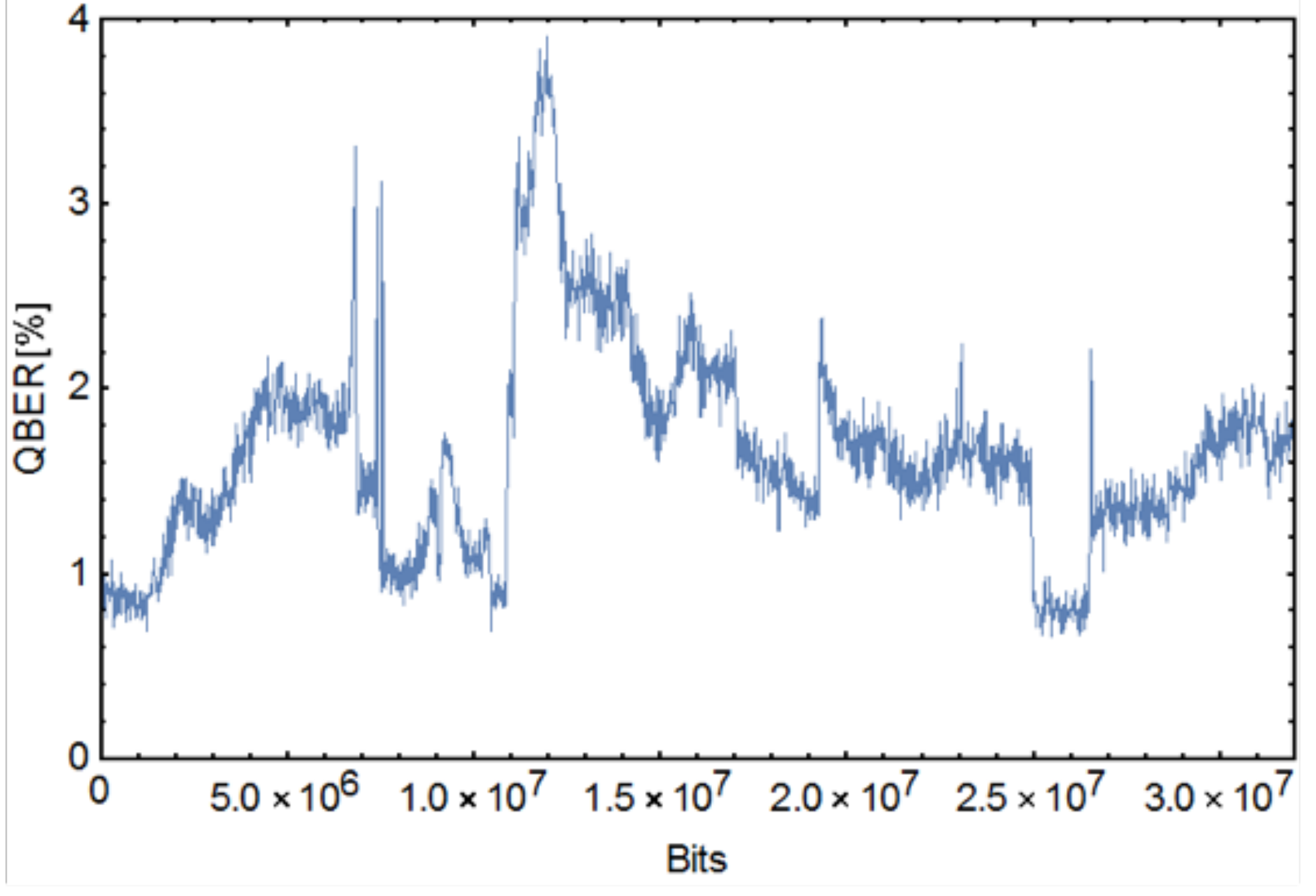}
\vskip -3mm
\caption
{QBER level is shown as function of distributed key size at five-day continuous experiment.}
\label{fig:modular}
\end{center}
\end{figure}

Fig.~3 demonstrates the QBER vs distributed key size, obtained as a result of a five-day continuous operation in the laboratory, including unsupervised operation on weekends. 
A truly random 70.2 MB key with 2.6\% QBER in average was amassed. 
Graph plotting is made with QBER averaging for every 2 KB of the distributed key. 
Because of the system's unstable operation, originating from both software and hardware issues, every several QKD sessions a failure occurred resulting in the very noisy behavior of QBER. 
After each failure the system restored and continued to work automatically. 
Smooth changes in QBER are mainly related with the time window adjustment shift because of temperature drift.
Later a stability of the system operation was achieved; some tests outside the laboratory were carried out when the system was on an exhibition. 
The system demonstrated much less noisy operation there, Fig.~4 shows the corresponding QBER. 
During the 18 hours of intermittent operation, a 4 MB key was distributed with 1.6\% average QBER. 
The abrupt steps of QBER are connected to changes of Alice's optic signal attenuation, the lowest regions correspond to approximately 0.2 photons/pulse.

With the fully functional plug\&play scheme based on our modular system, QKD tests were conducted between the two local branches of a bank in Moscow [5]. 
The quantum channel fiber length was 30.6 km, the total attenuation equal to 11.7 dB. 
During several hours, it was generated a quantum key at a speed of 1.0 kbit/s and an average QBER equal to 5.7\%, with an impulse intensity level of 0.18 photons. 
With an intensity of 0.35 photons exiting the sender, the corresponding parameters were equal to 1.9 kbit/s and 5.1\%. 
It was calculated the highest achievable speeds of key generation under these conditions to be equal to 1.1 and 2.2 kbit/s, correspondingly.

The software in charge of controlling the system was written in the development environment LabVIEW, created by National Instruments. 
It is a popular tool used by scientists and others, mostly for data acquisition, instrument control and industrial automation. 
We consider LabVIEW to be an optimal choice for educational purposes. 
It is already widely used in universities physical workshops, so it will be familiar to many students. 
At the same time, those who don't have any experience, can easily get involved, since the programming language is graphical and intuitive. 
Without diminishing its simplicity, it is also a very powerful tool --- many complex tasks usually associated with working with electronics are taken care of by the software itself. 
The FPGA boards used were chosen because of the simplicity with which they interact with the software. 
We should not forget to mention that post-processing work can also be carried out in the environment, since it offers all the needed instruments for data analysis.

Understanding how all the signals in the QKD device are related to each other and the electronic drivers that they operate can be a bit of a challenge. 
In order to solve rather complex problems concerning signal processing and synchronization, students will need to gain a deep understanding of visual dataflow programming, 
as well as getting used to what stands behind most of the graphical elements used. 
However, this challenge will allow students to better understand the subtleties of the software as well as the device itself.
Needless to say, the FPGA boards have their drawbacks. 
For example, the PCI-7811R board does not have the resources required for our device, and thus we had to use a third party FPGA to fulfil that function. 
The board does not have enough memory to delay information from a train of impulses by values of about 100-1000$\mu$s, which is needed to modulate the impulses by different devices correctly. 
Another reason for this problem is that at high frequencies the board can only deal with 8-bit integers, which would also mean that even if we could solve the memory problem, 
we would still be unable to change the delay with sufficiently precise increments. 
However, in the future we plan to use a more powerful version of the FPGA --- PCIe-7841R, 
which will have the required resources and allow us to implement all the needed functions solely on the board itself, without using any third party FPGA.

Using the presented desktop modular QKD system one is able to assemble much more complex schemes than the described one. 
For example, in some realizations an optic modulator is necessary to variate Alice's outgoing optic pulses. 
The proven possibility to attack such a scheme by means of bright illuminations to dazzle single photon detectors, demands the use of a flare detector on Bob's side for industrial QKD realizations. 
All these options can be included in our modular platform as it has enough room for specifications. 
As of this writing we are finalizing development of our compact single photon detectors that will be positioned under the optic table of Bob unit. 
As a result, we hope to significantly reduce the cost of the system.

The plug\&play scheme implementation we used allows to process optical pulses with a higher repetition frequency than we have done, 
in the near future we are going to increase the frequency at least up to 40 MHz. 
However, concerning the transfer rate, the plug\&play scheme has certain limitations because the optic pulses pass the quantum channel twice. 
From this point of view one-pass optic schemes with polarization encoding appear to be much more prospective, allowing much higher key distribution bitrate. 
But in one-pass fiber schemes polarization fluctuation problems arise, which need to be solved. 
Thus, there is a vast field for further QKD research and development. 
The presented modular QKD system has all necessary features such as flexibility, usability and robustness and can be aimed for both education and scientific research.

This research is carried out with the financial support of the state represented by the Ministry of Education and Science of the Russian Federation. 
Agreement 14.582.21.0009, ID RFMEFI58215X0009.

\end{document}